\documentclass[letters,fleqn,usenatbib]{mnras}

\usepackage{newtxtext,newtxmath}

\usepackage[T1]{fontenc}

\DeclareRobustCommand{\VAN}[3]{#2}
\let\VANthebibliography\thebibliography
\def\thebibliography{\DeclareRobustCommand{\VAN}[3]{##3}\VANthebibliography}

\usepackage{graphicx}	
\usepackage{amsmath}	
\usepackage{array,multirow,graphicx}
\usepackage{relsize}
\usepackage{makecell}
\usepackage{xcolor}
\usepackage[T1]{fontenc}

\title[Spectral Type and Geometric Albedo of 2001 CC21]{Spectral Type and Geometric Albedo of (98943) 2001 CC21, \\the {\it Hayabusa2\#} Mission Target}

\author[J. Geem et al.]{Jooyeon Geem,$^{1,2} $ \thanks{E-mail: ksky0422@snu.ac.kr} 
Masateru Ishiguro,$^{1,2} $ \thanks{E-mail: ishiguro@snu.ac.kr}
Mikael Granvik, $^{3,4}$ 
Hiroyuki Naito,$^{5}$ 
Hiroshi Akitaya,$^{6,7}$
\newauthor
Tomohiko Sekiguchi,$^{8}$
Sunao Hasegawa,$^{9}$ 
Daisuke Kuroda,$^{10}$ 
Tatsuharu Oono,$^{11}$
Yoonsoo P. Bach,$^{1,2}$
\newauthor
Sunho Jin,$^{1,2}$
Ryo Imazawa,$^{12}$
Koji S. Kawabata$^{7}$
Seiko Takagi,$^{13}$
Makoto Yoshikawa,$^{9}$
Anlaug A. Djupvik,$^{14,15}$
\newauthor
Julie Thiim Gadeberg,$^{14}$ 
Tapio Pursimo,$^{14,15}$
Oliver Durfeldt Pedros,$^{14,16}$ 
Jeppe Sinkbaek Thomsen,$^{14,17}$ 
\newauthor
Zuri Gray$^{14,18,19}$ 
\\
$^{1} $Department of Physics and Astronomy, Seoul National University, 1 Gwanak-ro, Gwanak-gu, Seoul 08826, Republic of Korea\\
$^{2} $SNU Astronomy Research Center, Seoul National University, 1 Gwanak-ro, Gwanak-gu, Seoul 08826, Republic of Korea\\
$^{3}$Department of Physics, University of Helsinki, PO. Box 64, FI-00014 Helsinki, Finland\\
$^{4}$ Asteroid Engineering Laboratory, Lule\aa\, University of Technology, Box 848, SE-98128 Kiruna, Sweden\\
$^{5} $Nayoro Observatory, 157-1 Nisshin, Nayoro, Hokkaido 096-0066, Japan\\
$^{6} $Planetary Exploration Research Center, Chiba Institute of Technology, 2-17-1 Tsudanuma, Narashino, Chiba 275-0016, Japan\\
$^{7} $Hiroshima Astrophysical Science Center, Hiroshima University, Kagamiyama 1-3-1, Higashi-Hiroshima, Hiroshima 739-8526, Japan\\
$^{8} $Asahikawa Campus, Hokkaido University of Education, 9 Hokumon, Asahikawa, Hokkaido 070-8621, Japan
\\
$^{9} $Institute of Space and Astronautical Science (ISAS), Japan Aerospace Exploration Agency (JAXA), Sagamihara, Kanagawa 252-5210, Japan\\
$^{10}$Bisei Spaceguard Center, Japan Spaceguard Association, 1716-3 Okura, Bisei-cho, Ibara, Okayama 714-1411, Japan\\
$^{11}$Department of Cosmosciences, Graduate School of Science, Hokkaido University, Kita-ku, Sapporo, Hokkaido 060-0810, Japan\\
$^{12}$Department of Physics, Graduate School of Advanced Science and Engineering, Hiroshima University
Kagamiyama, 1-3-1 Higashi-Hiroshima, \\
$\,\,\,\,$Hiroshima 739-8526, Japan\\
$^{13}$Department of Earth and Planetary Sciences, Faculty of Science, Hokkaido University, Kita-ku,
Sapporo, Hokkaido 060-0810, Japan\\
$^{14}$Nordic Optical Telescope, Rambla Jos\'{e} Ana Fern\'{a}ndez P\'{e}rez 7, ES-38711 Bre\~{n}a Baja, Spain\\
$^{15}$Department of Physics and Astronomy, Aarhus University, Ny Munkegade 120, DK-8000 Aarhus C, Denmark\\
$^{16}$DTU Space, National Space Institute, Technical University of Denmark,
Elektrovej 328, DK-2800 Kgs. Lyngby, Denmark\\
$^{17}$Dipartimento di Fisica e Astronomia, Universit{\fontencoding{T1}\selectfontà} di Bologna, Via Zamboni, 33 - 40126 Bologna, Italia\\
$^{18}$Armagh Observatory and Planetarium, College Hill, Armagh BT61 9DG, UK\\
$^{19}$Mullard Space Science Laboratory, Department of Space $\&$ Climate Physics, University College London, Holmbury St. Mary, Dorking, Surrey RH5 6NT, UK\\
}

\date{Accepted 20XX. Received 2023; in original form ZZZ}

\pubyear{2023}

\begin{document}
\label{firstpage}
\pagerange{\pageref{firstpage}--\pageref{lastpage}}
\maketitle

\begin{abstract}
We conducted optical polarimetry and near-infrared spectroscopy of JAXA's {\it Hayabusa2\#} mission target, (98943) 2001 CC21, in early 2023. Our new observations indicated that this asteroid has a polarimetric inversion angle of $\sim$21\degr, absorption bands around 0.9 and 1.9 $\mu$m, and a geometric albedo of 0.285$\pm$0.083. All these features are consistent with those of S-type but inconsistent with L-type. Based on this evidence, we conclude that JAXA's {\it Hayabusa2\#} spacecraft will explore an S-type asteroid with albedo and size (0.42--0.56 km when we assume the absolute magnitude of 18.6) similar to (25143) Itokawa.
\end{abstract}




\section{INTRODUCTION}

(98943) 2001 CC21 (hereafter CC21) is the target of the {\it Hayabusa2} extended mission (the nickname is {\it Hayabusa2\#}) operated by the Japan Aerospace Exploration Agency (JAXA). After a successful sample return from (162173) Ryugu in 2020, {\it Hayabusa2} spacecraft plans to explore its next targets, CC21 in July 2026 and 1998 KY26 in July 2031. 

CC21 has a rotation period of 5 hours \citep{Hirabayashi+2021}. However, little is known about CC21. Initially, \citet{Binzel+2004} reported that CC21 is classified as L-type. On the contrary, \citet{Lazzarin+2005} and \citet{DeMeo+2009} pointed out the possibility of an S-complex asteroid (either Sk- or Sw-type). Therefore, the taxonomic type of this space mission target still needs to be examined more thoroughly. In addition, it should be emphasized that CC21's albedo has yet to be determined. Knowing albedo is critical for setting appropriate exposure times during the {\it Hayabusa2\#}'s fast flyby and estimating the size.

We present our new observational evidence for the taxonomic type and the geometric albedo of this asteroid, taking advantage of the observation opportunity in 2023 January--March. We conducted optical polarimetry in a wide range of phase angles ($\alpha=21.7\degr$--$99.9\degr$), which allows for estimating the taxonomic classification and the geometric albedo. Moreover, we conducted the near-infrared (NIR, $0.7$--$2.4\,\mu$m) spectroscopy at an intermediate phase angle ($\alpha\sim32\degr$). We describe our observations and data reduction in Section \ref{sec:observations} and the results in section \ref{sec:Result}. Based on the results, we discuss our results in section \ref{sec:discussion}.

\section{Observations and data analysis}
\label{sec:observations}
\subsection{Optical Polarimetry}
\label{sec:method_pol}

The observation circumstances are summarized in Table \ref{table:obs&result}. We obtained the polarimetric data of CC21 by using three instruments: the FAPOL polarimeter of the Alhambra faint object spectrograph and camera (ALFOSC) on the 2.56-m Nordic Optical Telescope at the Observatorio del Roque de los Muchachos, the Hiroshima optical and near-infrared camera (HONIR; \citealt{Akitaya+2014}) on the 1.5-m Kanata Telescope at the Higashi-Hiroshima observatory and the visible multi-spectral imager (MSI; \citealt{Watanabe+2012}) on the 1.6-m Pirka Telescope at the Nayoro Observatory of Hokkaido University. We employed $V$- and $R$-band. 
All instruments are composed of a rotatable half-wave plate (HWP) and mounted in the Cassegrain focus of each telescope. HONIR and MSI equip a Wollaston prism, and ALFOSC employs a calcite plate mounted in the aperture wheel. To obtain sets of the Stokes parameters, we rotated the HWP in the order of $0\,\degr$, $45\,\degr$, $22.5\,\degr$ and $67.5\,\degr$ for HONIR and MSI and from $0\,\degr$ to $337.5\,\degr$ at $22.5\,\degr$ interval for ALFOSC.

\begin{table*}
\centering
\caption{Summary of Polarimetry}
\label{table:obs&result}
\begin{tabular}{lcccccccccccccc}
\hline
Date in UT$^a$  & Inst$^b$ & Filter & Exp$^c$ &N$^d$ &  $ r^e $ & $ \Delta^f $ &  $ \phi^g $& $ \alpha^h$&$P^i$&$\sigma \,P^j$&$\theta_{P}^k$&$\sigma \, \theta_{P}^l$ &${P_\mathrm{r}}^m$&${\theta_\mathrm{r}}^n$ \\
 &  && (s)& & ($ \mathrm{au} $) & ($ \mathrm{au} $) & ($\degr$) & ($\degr$)&($\%$)&($\%$)&($\degr$)&($\degr$)&($\%$)&($\degr$)\\
   \hline
 Jan 24 18:35--18:49 &  HONI$R$ &             $R$ &          60 &       12 &  1.14 &   0.17 &  213.0 &   21.7 &  0.38 &  0.30 &     83.7 &      22.1 &  0.18 &    -31.3 \\
 Jan 25 17:40--18:00 &  HONI$R$ &             $R$ &          60 &       16 &  1.14 &   0.17 &  208.5 &   22.1 &  0.00 &  0.19 &     31.7 &      52.0 & -0.00 &     45.2 \\
 Jan 29 14:35--15:14 &    MSI &            $R$ &         120 &       20 &  1.13 &   0.16 &  191.2 &   24.5 &  0.31 &  0.19 &    -85.0 &      17.3 &  0.30 &     -8.1 \\
 Jan 29 15:34--16:45 &    MSI &             V &         150 &       20 &  1.13 &   0.16 &  190.9 &   24.5 &  0.45 &  0.25 &    -89.8 &      16.2 &  0.37 &    -16.7 \\
 Feb 03 23:01--23:10 &  FAPOL &  $V$ &         120 &        8 &  1.11 &   0.15 &  170.3 &   30.4 &  1.25 &  1.15 &     54.4 &      26.3 &  0.75 &    -26.6 \\
 Feb 09 13:39--16:18 &    MSI &            $R$ &         120 &       16 &  1.09 &   0.14 &  151.5 &   38.6 &  1.22 &  0.53 &     63.5 &      12.4 &  1.19 &     -6.8 \\
 Feb 09 16:32--16:50 &    MSI &             V &         180 &        8 &  1.09 &   0.14 &  151.3 &   38.7 &  1.62 &  0.73 &     73.5 &      12.9 &  1.62 &     -2.8 \\
 Feb 11 23:30--23:47 &  FAPOL &  $V$ &         120 &       12 &  1.09 &   0.14 &  144.3 &   42.4 &  1.89 &  0.43 &     49.8 &       6.4 &  1.87 &     -4.5 \\
 Feb 15 13:05--13:28 &    MSI &            $R$ &          90 &        8 &  1.07 &   0.14 &  133.4 &   48.4 &  2.09 &  0.50 &     45.8 &       6.9 &  2.06 &     -5.2 \\
 Feb 15 13:55--14:29 &    MSI &             V &     120,240 &     12,4 &  1.07 &   0.14 &  133.3 &   48.5 &  2.92 &  0.56 &     35.9 &       5.5 &  2.85 &      6.0 \\
 Feb 17 12:03--14:52 &    MSI &            $R$ &      90,120 &    32,24 &  1.07 &   0.14 &  127.4 &   51.9 &  2.53 &  0.12 &     38.0 &       1.4 &  2.53 &      0.9 \\
 Feb 17 13:06--16:08 &    MSI &             V &  90,120,150 &  4,24,24 &  1.07 &   0.14 &  127.2 &   52.0 &  2.76 &  0.17 &     36.9 &       1.8 &  2.75 &     -1.8 \\
 Feb 26 20:42--20:59 &  FAPOL &  $V$ &         120 &       12 &  1.03 &   0.13 &  100.1 &   68.0 &  4.13 &  0.19 &      8.8 &       1.3 &  4.13 &     -1.3 \\
 Feb 26 16:40--16:53 &    MSI &            $R$ &         180 &        8 &  1.03 &   0.13 &  100.5 &   67.7 &  3.28 &  1.18 &      5.5 &      10.3 &  3.24 &     -4.7 \\
 Feb 26 18:28--19:28 &    MSI &             V &         240 &       12 &  1.03 &   0.13 &  100.3 &   67.9 &  4.61 &  1.16 &     22.4 &       7.2 &  4.60 &     -0.9 \\
 Mar 04 14:15--15:20 &    MSI &            $R$ &         180 &       16 &  1.01 &   0.13 &   84.6 &   78.0 &  4.66 &  0.88 &      8.4 &       5.4 &  4.64 &     -2.5 \\
 Mar 04 11:56--13:48 &    MSI &             V &         180 &       20 &  1.01 &   0.13 &   84.9 &   77.8 &  5.51 &  1.33 &     -6.5 &       6.9 &  5.50 &      1.7 \\
 Mar 05 20:35--20:43 &  FAPOL &  $V$ &         120 &       16 &  1.01 &   0.13 &   81.5 &   80.2 &  5.69 &  0.39 &     -7.7 &       2.0 &  5.69 &      0.8 \\
 Mar 13 21:07--21:07 &  FAPOL &  $V$ &         120 &        4 &  0.98 &   0.13 &   62.6 &   94.4 &  6.02 &  0.48 &    -27.8 &       2.3 &  6.01 &     -0.5 \\
 Mar 16 20:24--20:41 &  FAPOL &  $V$ &         120 &       12 &  0.96 &   0.13 &   56.1 &   99.9 &  6.33 &  0.63 &    -33.5 &       2.9 &  6.22 &     -5.4 \\
\hline
\multicolumn{15}{l}{$ ^a $ UT at exposure start,$ ^b $ Instrument, $ ^c $Exposure time, $ ^d $ Number of valid images, $^e$ Median heliocentric distance, $ ^f $ Median geocentric}\\
\multicolumn{15}{l}{ distance, $ ^g $ Position angle of the scattering plane, $ ^h $ Median solar phase angle, $^i$ Nightly averaged polarization degree, $^j$ Uncertainty of $P$, $ ^k $ Position }\\
\multicolumn{15}{l}{angle of the strongest electric vector, $^l$ Uncertainty of $\theta_\mathrm{P}$, $^m$ Polarization degree referring to the scattering plane, $^n$ Position angle referring to the scattering}\\
\multicolumn{15}{l}{  plane.}\\
\multicolumn{14}{l}{The web-based JPL Horizon system (\url{http://ssd.jpl.nasa.gov/?horizons}) was used to obtain $r$, $ \Delta$, $ \phi$, and $ \alpha$ in the table.}\\
\end{tabular}
\end{table*}

\begin{table*}
\centering
\caption{Calibration Parameters of Polarimetry}
\label{table:calibration}
\begin{tabular}{lcccccccc}
\hline
Inst&Filter&$P_\mathrm{eff}^{a}$ & $q_\mathrm{inst}$ & $u_\mathrm{inst}$ & $\theta_\mathrm{off}$& UP$ ^b $  & SP$ ^c $&Ref$^{d}$ \\
&&($\%$)& ($\%$)&($\%$)&($\degr$)&&&\\
\hline
HONIR & $R$ & $97.58$ & $0.010\pm0.050$ & $-0.008\pm0.037$  & $38.00\pm0.83$
&  G191B2B  &  HD 29333, HD 251204&(1),(2)\\
ALFOSC & $V$& 100 (assumed) & $0.012\pm0.065$ & $-0.054\pm0.055$ & $-87.78\pm0.10$ 
&  HD 42182, HD 65629 &  BD+59 389& (3)\\
MSI & $V$ & $99.59\pm0.02$ & $0.785 \pm 0.020$ & $1.077 \pm 0.019$ &  $-20.61\pm 0.26$
& HD 15318&HD 7927&(4)\\
&$R$&$99.55\pm 0.01$ & $0.584\pm 0.011$ & $0.751 \pm 0.011$ & $-17.09\pm 0.52$
& HD 15318&HD 7927&(4)\\
\hline
\multicolumn{9}{l}{$ ^a $ the polarization efficiency,$ ^b $ Unpolarized standard stars, $ ^c $ Strongly polarized standard stars, $ ^d $ the references of standard stars }\\
\multicolumn{9}{l}{(1) \citet{Turnshek+1990}, (2)\citet{Whittet+1992}, (3)\citet{Schmidt+1992}, (4)\citet{Wolff+1996}}\\
\end{tabular}
\end{table*}

 The polarimetric data were analyzed in the same manner as described in \citet{Ishiguro+2022} for MSI data and in \citet{Geem+2022} for ALFOSC and HONIR data. We observed unpolarized standard stars to obtain the instrumental polarization parameters ($q_\mathrm{inst}$ and $u_\mathrm{inst}$) and strongly polarized stars to determine the position angle offset ($\theta_\mathrm{off}\equiv\theta_\mathrm{cat}-\theta_\mathrm{obs}$). Here, $\theta_\mathrm{cat}$ and $\theta_\mathrm{obs}$ are the position angle of a star from a catalog and an observation, respectively. 
We summarize the instrument calibration parameters in Table \ref{table:calibration} and the derived polarimetric degrees in Table \ref{table:obs&result}. 
In Table \ref{table:obs&result}, the polarization degree with respect to the scattering plane ($P_\mathrm{r}$) is given, as is conventionally driven for asteroid polarimetry. We fit the polarization phase curve (PPC) by using the Lumme--Muinonen function (L/M, \citealt{Lumme+1993}) and the linear function by employing the Markov chain Monte Carlo method implemented in PyMC3 \citep{Salvatier+2016}. 
10,000 samples per chain with four chains are adopted. We used the same boundary conditions to derive the uncertainties of the optimal parameters written in \citet{Geem+2022}. The initial guesses of each parameter are ($h$, $\alpha_{0}$, $c_1$, $c_2$) = ($0.07\,\mathrm{percent}\,\mathrm{cent}\,\mathrm{deg}^{-1}$, $20\degr$, 0.1,  0.001). Beyond the $\alpha_{0}$, it is known that $P_\mathrm{r}$ of intermediate or high-albedo asteroids, such as S-complex asteroids, pseudo-linearly increases with increasing $\alpha$ up to the maximum phase angle \citep{Cellino+2005}. Thus, we applied the linear function to derive $h$ and $\alpha_{0}$ by using the data at $\alpha<80\degr$. The best-fitting results and their uncertainties obtained with the two different fitting functions agree with each other. The results obtained using L/M cover those obtained by the linear fitting. Therefore, we discuss only the results derived from L/M hereafter.

\subsection{Near-infrared Spectroscopy}
\label{sec:method_spec}
The NIR spectral data (0.7--$2.4\,\mu$m) were obtained during two nights, 2023 February 5--6 UT, using the SpeX instrument at the Mauna Kea Observatory 3.2-m NASA Infrared Telescope Facility (IRTF). We used the 0.8 arcsec width slit aligned with the parallactic angle \citep{Binzel+2019}. We obtained 86 individual asteroidal spectral images with an integration time of 120s each. The observational circumstances are summarized in Table \ref{table:spec obs}. The standard star of SAO 42382 (G2V) was observed, which is closely located from the asteroid (the airmass difference$<$0.1). We used Spextool, an \textsc{IDL}-based spectral reduction program \footnote{\url{http://irtfweb.ifa.hawaii.edu/research/ dr_resources/}} for data reduction (i.e., flat-fielding, sky subtraction, spectrum extraction, and wavelength calibration.).

\begin{table}
\centering
\caption{Observation Circumstance of Spectroscopy}
\label{table:spec obs}
\begin{tabular}{lccccc}
\hline
Date in UT$^a$  & Exp & $ \alpha^b$ & Airmass &N$^c$& Solar Analog  \\
               & (s) &($\degr$)  & & & \\
   \hline
Feb 05 09:48--12:39 & 120 & 32.3 & 1.18--1.41 & 38 &SAO 42382\\
Feb 06 06:08--09:38 & 120 & 33.7 & 1.20--1.69 & 48 &SAO 42382\\
\hline
\multicolumn{6}{l}{$ ^a $ UT at exposure start, $ ^b $solar phase angle, $^c$ number of valid images}\\
\end{tabular}
\end{table}

\section{Results}
\label{sec:Result}

\begin{figure}
	\includegraphics[width=\columnwidth]{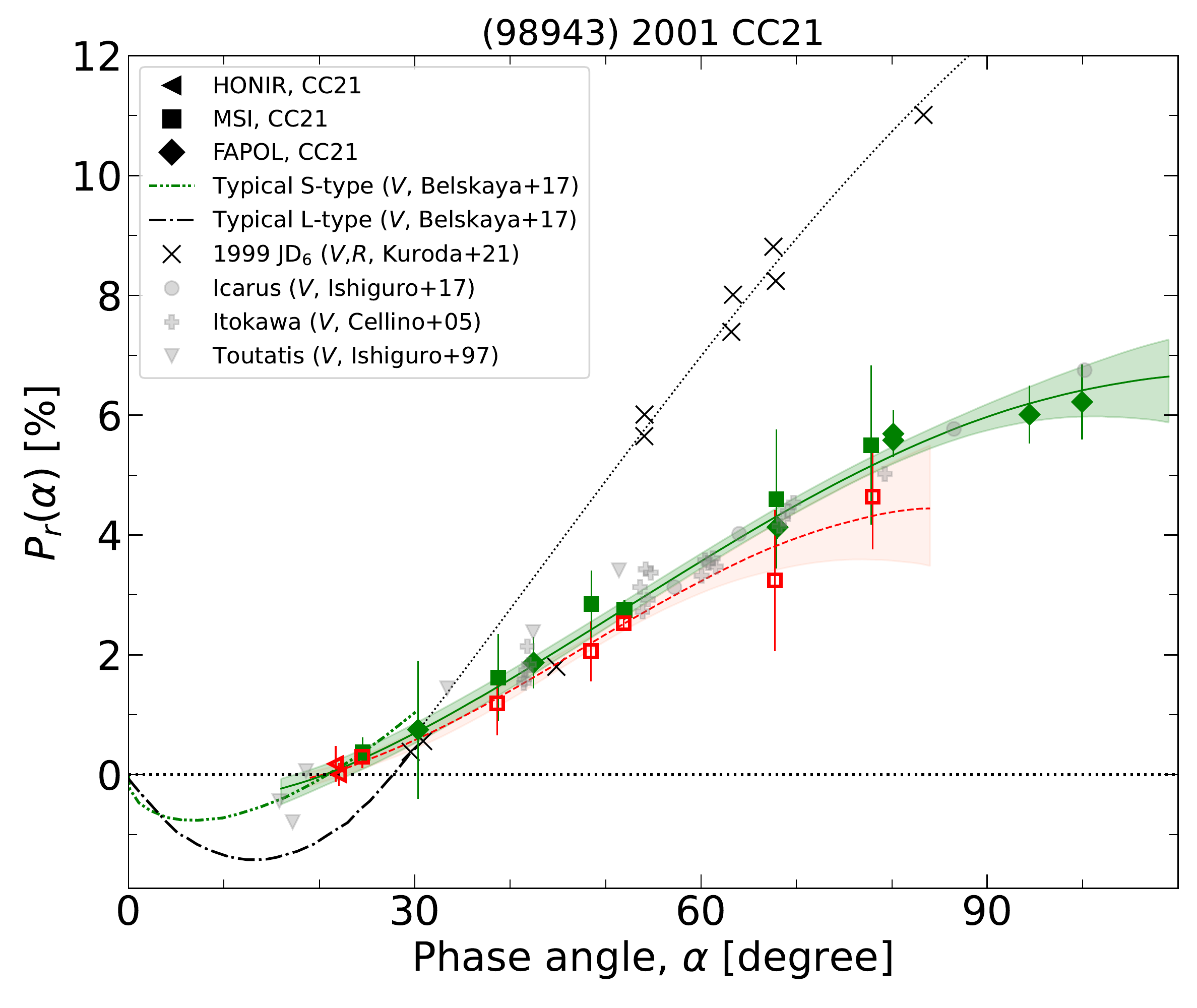}
    \caption{Phase angle ($\alpha$) dependence of polarization degree ($P_\mathrm{r}$). Data taken by HONIR, MSI, and FAPOL are shown by the triangle, square, and diamond markers, respectively. The filled and empty markers indicate $V$- and $R$-bands. The solid and dash lines are curves that fit data in $V$- and $R$-bands, respectively, by using the L/M function. The dotted line is the PPC of 1999 JD$_6$.
    }
    \label{fig:phase-plot}
\end{figure}

Figure \ref{fig:phase-plot} indicates the nightly-averaged PPC. Similar to other asteroids, it shows an upward trend in the observed phase angles, having an inversion angle ($\alpha_0$) of around 20$\,\degr$ and a maximum polarization phase ($\alpha_\mathrm{max}$) of around 100$\,\degr$. Table \ref{table:fitting} summarizes these key parameters in PPC. For comparison, we plot the typical PPC of the S-type and L-type in low phase angles \citep{Belskaya+2017}. We also compare CC21 with S-type asteroids: (1566) Icarus \citep{Ishiguro+2017}, (4179) Toutatis \citep{Ishiguro+1997}, and (25143) Itokawa \citep{Cellino+2005}, in $V$-band and with an L-type asteroid (85989) $1999\, \mathrm{JD}_\mathrm{6}$ in $V$- and $R$- bands \citep{Kuroda+2021}. A glance at Figure \ref{fig:phase-plot} finds that CC21 is closer to these S-type than to the L-type.

\begin{table}
\centering
\caption{PPC Fitting Result}
\label{table:fitting}
\begin{tabular}{lcccc}

\hline
Filter& slope $h$ & $\alpha_{0}$ & $P_\mathrm{max}$ & $\alpha_\mathrm{max}$\\
& ($\%\,$deg$^{-1}$) &($\degr$)  & ($\%$) & ($\degr$) \\
   \hline
$V$ & $0.07^{+0.02}_{-0.02}$& $20.7^{+3.3}_{-2.6}$ & $6.7^{+1.0}_{-0.6}$ & $ 114.5^{+5.4}_{-12.2}$\\
$R$  & $0.06^{+0.01}_{-0.02}$& $21.2^{+2.2}_{-2.2}$ & $4.3^{+3.5}_{-1.2}$ & $ 88.4^{+31.5}_{-8.4}$\\
\hline
\end{tabular}
\end{table}

It is known that the polarimetric slope $h$ is a good proxy for $p_\mathrm{V}$. The relation between $h$ and $p_\mathrm{V}$ is given by $\log_{10} \left( p_\mathrm{V} \right) = C_1 \log_{10} \left( h \right) + C_2 $, where $ C_\mathrm{1} $ and $C_\mathrm{2} $ are constants \citep{Geake+1986}. We derived the $p_\mathrm{V}$ by substituting $h= 0.07 \pm 0.02\,\mathrm{per}\,\mathrm{cent}\,\mathrm{deg}^{-1}$ in $V$-band and obtained $p_\mathrm{V}=$  $0.285 \pm 0.083$ and $p_\mathrm{V}=$ $0.284 \pm 0.076$ using $C_1$ and $C_2$ values in \citet{Lupishko+2018} and \citet{Cellino+2015} (for $p_\mathrm{V}>0.08$), respectively. Although these two albedo estimates are very close, we will henceforth use the former value ($p_\mathrm{V}=$  $0.285 \pm 0.083$), which has the larger error, for safety.

\begin{figure}
	\includegraphics[width=\columnwidth]{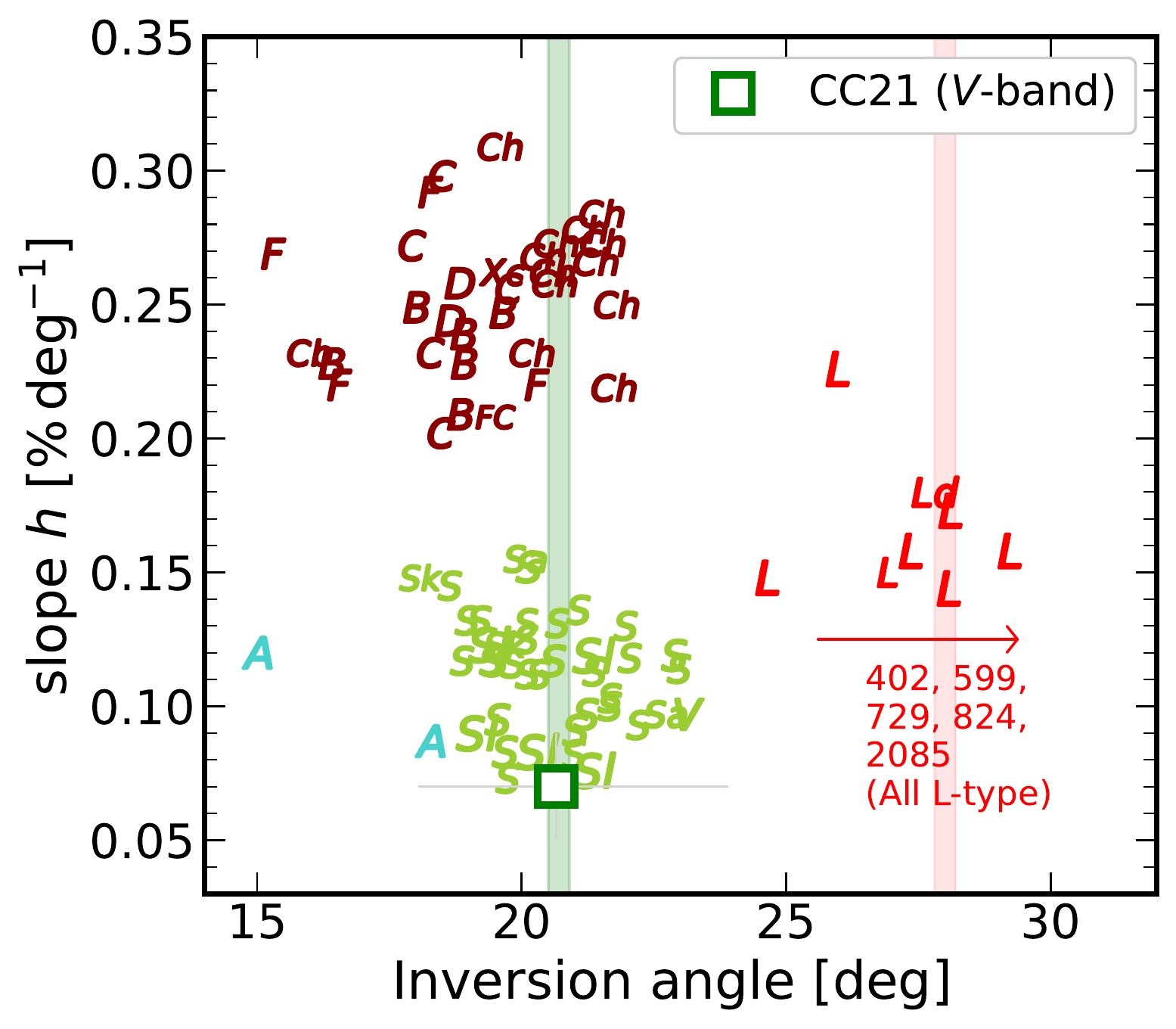}
    \caption{A comparison with other types of asteroids \citep{Kuroda+2021, Lupishko+2022}. Each marker indicates the taxonomic type of asteroid (Tholen or Bus--DeMeo type). CC21 in $V$-band is shown as the green empty box. L-type asteroids with  $\alpha_{0} \gtrsim25\degr$ and unknown $h$ are shown below the arrow \citep{Lupishko+2022}. The green and red colored boxes show the typical $\alpha_{0}$ ranges of the S-type and the L-type, respectively, from \citet{Belskaya+2017}.}
    \label{fig:comparison}
\end{figure}

We created $\alpha_{0}$--$h$ plot to discriminate the polarization properties of L-, S-, and other types of asteroids using databases in \citet{Kuroda+2021} and \citet{Lupishko+2022} (Figure \ref{fig:comparison}). Because $h$ and $p_\mathrm{V}$ are inversely correlated, asteroids with lower $p_\mathrm{V}$ are typically found higher up in the plot. L-type asteroids are known to have distinctively larger $\alpha_0$ than other asteroids. From this comparison, $\alpha_{0}$--$h$ of CC21 matches those of S-types rather than L-types.

\begin{figure}
	\includegraphics[width=\columnwidth]{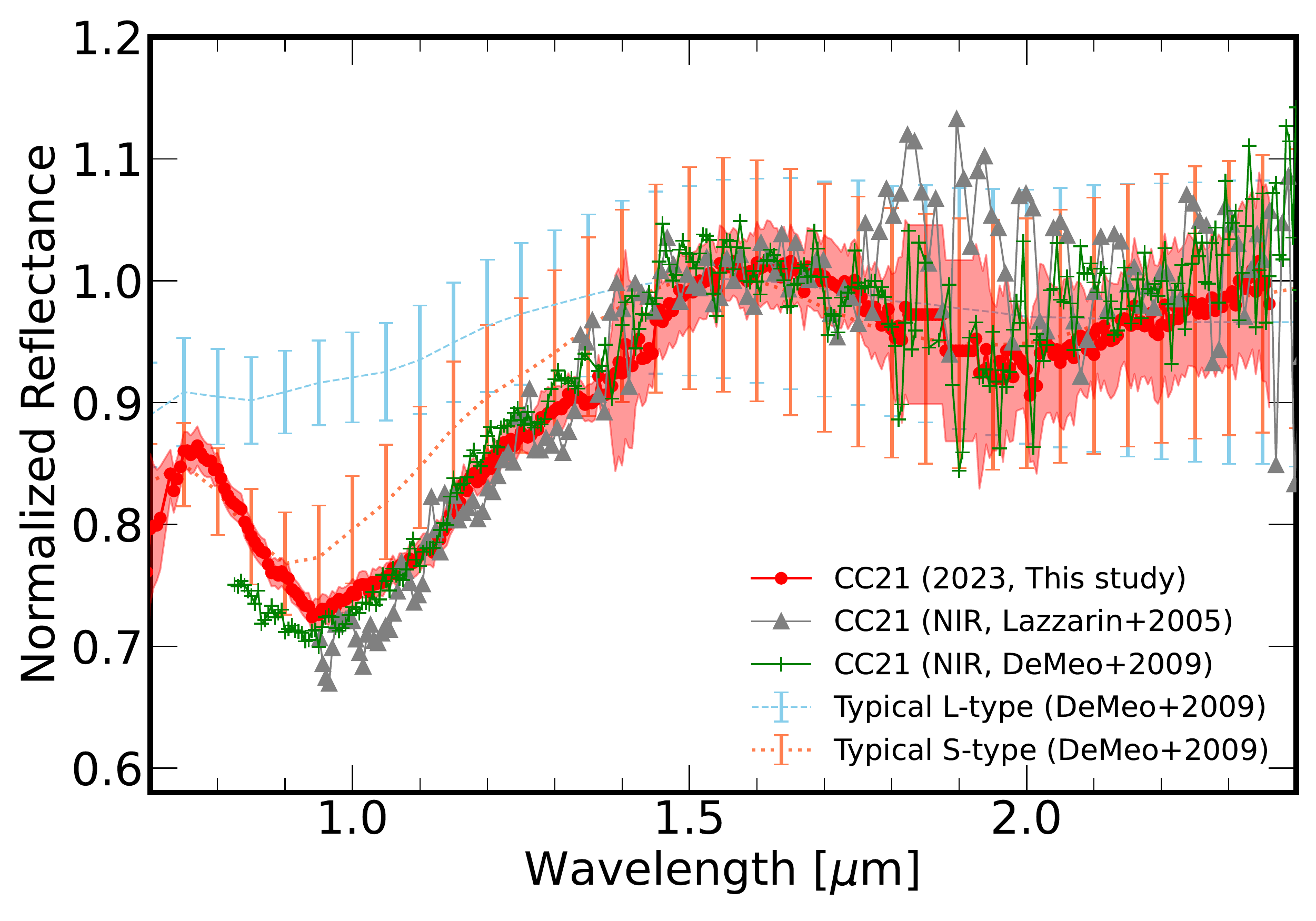}
    \caption{The combined spectra of CC21 from two nights' observations is colored red. For the comparison, NIR spectra of CC21 from the previous studies \citep{Lazzarin+2005, DeMeo+2009} by the gray triangle, green cross markets. The coral dotted and blue dashed lines show the typical spectra of S-type and L-type asteroids \citep{DeMeo+2009}, respectively. The reflectances are normalized at $1.6\,\mu$m. }
    \label{fig:spec_fig}
\end{figure}

The NIR spectra obtained over the two nights are consistent with each other. No rotational variation of the target spectra is found during the observations. For this reason, we combined all spectral data from two nights. Figure \ref{fig:spec_fig} shows the resultant spectrum.  
The plot compares CC21's reflectance spectrum with typical S- and L-type asteroids from \citet{DeMeo+2009}. Our CC21 spectrum indicates not only the clear absorption around $0.9\,\mu$m but also the shallow absorption around $1.9\,\mu$m associated with pyroxene. The continuum in the visible range at $0.75\,\mu$m is confirmed. These features are characteristic of S-type asteroids and are not found in L-type.

\section{Discussion}
\label{sec:discussion}

In the previous studies, CC21 was classified in either L- or S-type \citep{Binzel+2004,Lazzarin+2005,DeMeo+2009}. Our observations indicates S-type features and rules out L-type possibility. Although empirical, L-type asteroids are known to have distinctive $\alpha_0$ values compared to other asteroids. The $\alpha_0$ value in this study is predominantly different from that of L-type asteroids but is consistent with S-type asteroids. Another important aspect of our polarimetry is that the geometric albedo was determined. The derived CC21's albedo ($p_\mathrm{V}=0.285\pm0.083$) is in the range of S-type asteroids ($p_\mathrm{V}=0.258\pm0.087$, \citealt{DeMeo+2013}) but is slightly higher than the average L-type asteroids \citep[$p_\mathrm{V}=0.183 \pm 0.089$,][]{DeMeo+2013}. We also emphasize that our NIR spectrum with a good S/N provides definitive evidence for the taxonomic type. Our NIR spectrum agrees with the previous CC21's NIR spectra \citep{Lazzarin+2005, DeMeo+2009} within observational uncertainty \citep{Marsset+2020} (Figure \ref{fig:spec_fig}). The $0.9$ and $1.9\,\mu$m absorption features (typical of S-type) are detected. Using an online-based classification tool\footnote{\url{http://smass.mit.edu/cgi-bin/busdemeoclass-cgi}}, CC21 is further subcategorized into Sq or Q-type. These taxonomic types (less weathered ordinary chondritic asteroids) are common in the near-Earth region. Given all these observational results, we conclude that CC21 is an S-type rather than L-type. We expect that a discussion of the scientific case of  {\it Hayabusa2\#} will be conducted with the S-type in consideration.

The second significant point of our study is to determine the albedo. Because  {\it Hayabusa2\#} will conduct a fast flyby with CC21, the exposure times should be pre-determined based on the geometric albedo and scattering phase function. The geometric albedo of an S-type asteroid, (25143) Itokawa, was estimated by polarimetry before the launch \citep{Cellino+2005}. Notably, the geometric albedo estimated by the polarization method matches the actual value derived by the in-situ observation \citep{Lee2018,Tatsumi2018}. Considering that CC21 is an S-type asteroid, our albedo value obtained by the same method as \citet{Cellino+2005} is also quite reliable. Our albedo estimate is higher than the assumed value in \citet{Hirabayashi+2021} ($p_\mathrm{V}=0.15$). Although we need to refer to the absolute magnitude $H_\mathrm{V}=$18.6 from the literature \citep{Binzel+2004}, the effective diameter of CC21 estimated from our albedo would be 0.42--0.56 km. Note that the size becomes smaller than the value in the previous research \citet{Hirabayashi+2021}.

Polarimetry at large phase angles is also useful for estimating particle size on the asteroid's surface. It is known that the $P_\mathrm{max}$ is correlated with the $p_\mathrm{V}$ and the grain size. Their relationship is given by $d = 0.03\exp{2.9(\log{(10^2 A)} + 0.845\log{(10P_\mathrm{max})})}$, where $d$ is the grain size in $\mu$m and $A$ is an albedo at $\alpha = 5\degr$ \citep{Shkuratov+1992}. We calculate CC21's $A=$$0.198\pm 0.058$ by considering the intensity ratio of $I(0\fdg3)/I(5\degr) = 1.44\pm0.04$ for typical S-type asteroids \citep{Belskaya+2000}. As a result, CC21 would be covered by grain with the size of $100$--$130\,\mu$m, like (1566) Icarus, another near-Earth S-type asteroid whose size is similar to CC21 ($\lesssim$ $1\,\mathrm{km}$). 

\section{Summary}
We conducted optical polarimetric and near-infrared spectroscopic observations of 2001 CC21 in early 2023. The inversion angle ($\alpha_{0}\sim21\degr$) and geometric albedo ($p_\mathrm{V}\sim 0.3$) are consistent with those of S-types but significantly different from those of L-types. The near-infrared spectrum of the target shows clear absorption bands around $0.9$ and $1.9\,\mu$m, which are the typical spectral features of S-type asteroids. Based on the results, we conclude that 2001 CC21 is a near-Earth S-type asteroid with an albedo of $p_\mathrm{V}=0.285\pm0.083$ and a size of 0.42--0.56 km.

\section*{Acknowledgements}
This research at SNU was supported by a National Research Foundation of Korea (NRF) grant funded by the Korean government (MEST) (No. 2023R1A2C1006180). SH was supported by the Hypervelocity Impact Facility (former name: The Space Plasma Laboratory), ISAS, JAXA. 
The data presented here were obtained with ALFOSC, which is provided by the Instituto de Astrofisica de Andalucia (IAA) under a joint agreement with the University of Copenhagen and NOT.
Based on observations made with the Nordic Optical Telescope, owned in collaboration by the University of Turku and Aarhus University, and operated jointly by Aarhus University, the University of Turku and the University of Oslo, representing Denmark, Finland and Norway, the University of Iceland and Stockholm University at the Observatorio del Roque de los Muchachos, La Palma, Spain, of the Instituto de Astrofisica de Canarias.

 \section*{Data Availability}\label{dataave}
The observational data are available in Zenodo\footnote{\url{https://doi.org/10.5281/zenodo.XXXXX}}. The source codes and scripts for the data analyses, plots, and resultant data tables are available via the GitHub service\footnote{\url{https://github.com/Geemjy/XXXXX.git}}.



\bibliographystyle{mnras}
\bibliography{reference.bib} 

\begin{thebibliography}{}
\makeatletter
\relax
\def\mn@urlcharsother{\let\do\@makeother \do\$\do\&\do\#\do\^\do\_\do\%\do\~}
\def\mn@doi{\begingroup\mn@urlcharsother \@ifnextchar [ {\mn@doi@}
  {\mn@doi@[]}}
\def\mn@doi@[#1]#2{\def\@tempa{#1}\ifx\@tempa\@empty \href
  {http://dx.doi.org/#2} {doi:#2}\else \href {http://dx.doi.org/#2} {#1}\fi
  \endgroup}
\def\mn@eprint#1#2{\mn@eprint@#1:#2::\@nil}
\def\mn@eprint@arXiv#1{\href {http://arxiv.org/abs/#1} {{\tt arXiv:#1}}}
\def\mn@eprint@dblp#1{\href {http://dblp.uni-trier.de/rec/bibtex/#1.xml}
  {dblp:#1}}
\def\mn@eprint@#1:#2:#3:#4\@nil{\def\@tempa {#1}\def\@tempb {#2}\def\@tempc
  {#3}\ifx \@tempc \@empty \let \@tempc \@tempb \let \@tempb \@tempa \fi \ifx
  \@tempb \@empty \def\@tempb {arXiv}\fi \@ifundefined
  {mn@eprint@\@tempb}{\@tempb:\@tempc}{\expandafter \expandafter \csname
  mn@eprint@\@tempb\endcsname \expandafter{\@tempc}}}

\bibitem[\protect\citeauthoryear{{Akitaya} et~al.,}{{Akitaya}
  et~al.}{2014}]{Akitaya+2014}
{Akitaya} H.,  et~al., 2014, \mn@doi [\SPIE] {10.1117/12.2054577}, \href
  {https://ui.adsabs.harvard.edu/abs/2014SPIE.9147E..4OA} {9147, 91474O}

\bibitem[\protect\citeauthoryear{{Belskaya} \& {Shevchenko}}{{Belskaya} \&
  {Shevchenko}}{2000}]{Belskaya+2000}
{Belskaya} I.~N.,  {Shevchenko} V.~G.,  2000, \mn@doi [\icarus]
  {10.1006/icar.2000.6410}, \href
  {https://ui.adsabs.harvard.edu/abs/2000Icar..147...94B} {147, 94}

\bibitem[\protect\citeauthoryear{{Belskaya} et~al.,}{{Belskaya}
  et~al.}{2017}]{Belskaya+2017}
{Belskaya} I.~N.,  et~al., 2017, \mn@doi [\icarus]
  {10.1016/j.icarus.2016.11.003}, \href
  {https://ui.adsabs.harvard.edu/abs/2017Icar..284...30B} {284, 30}

\bibitem[\protect\citeauthoryear{{Binzel}, {Perozzi}, {Rivkin}, {Rossi},
  {Harris}, {Bus}, {Valsecchi}  \& {Slivan}}{{Binzel}
  et~al.}{2004}]{Binzel+2004}
{Binzel} R.~P.,  {Perozzi} E.,  {Rivkin} A.~S.,  {Rossi} A.,  {Harris} A.~W.,
  {Bus} S.~J.,  {Valsecchi} G.~B.,   {Slivan} S.~M.,  2004, \mn@doi [\maps]
  {10.1111/j.1945-5100.2004.tb00098.x}, \href
  {https://ui.adsabs.harvard.edu/abs/2004M&PS...39..351B} {39, 351}

\bibitem[\protect\citeauthoryear{{Binzel} et~al.,}{{Binzel}
  et~al.}{2019}]{Binzel+2019}
{Binzel} R.~P.,  et~al., 2019, \mn@doi [\icarus]
  {10.1016/j.icarus.2018.12.035}, \href
  {https://ui.adsabs.harvard.edu/abs/2019Icar..324...41B} {324, 41}

\bibitem[\protect\citeauthoryear{Cellino, Yoshida, Anderlucci, Bendjoya, {Di
  Martino}, Ishiguro, Nakamura  \& Saito}{Cellino et~al.}{2005}]{Cellino+2005}
Cellino A.,  Yoshida F.,  Anderlucci E.,  Bendjoya P.,  {Di Martino} M.,
  Ishiguro M.,  Nakamura A.~M.,   Saito J.,  2005, \mn@doi [Icarus]
  {https://doi.org/10.1016/j.icarus.2005.08.005}, 179, 297

\bibitem[\protect\citeauthoryear{{Cellino}, {Bagnulo}, {Gil-Hutton}, {Tanga},
  {Ca{\~n}ada-Assandri}  \& {Tedesco}}{{Cellino} et~al.}{2015}]{Cellino+2015}
{Cellino} A.,  {Bagnulo} S.,  {Gil-Hutton} R.,  {Tanga} P.,
  {Ca{\~n}ada-Assandri} M.,   {Tedesco} E.~F.,  2015, \mn@doi [\mnras]
  {10.1093/mnras/stv1188}, \href
  {https://ui.adsabs.harvard.edu/abs/2015MNRAS.451.3473C} {451, 3473}

\bibitem[\protect\citeauthoryear{{DeMeo} \& {Carry}}{{DeMeo} \&
  {Carry}}{2013}]{DeMeo+2013}
{DeMeo} F.~E.,  {Carry} B.,  2013, \mn@doi [\icarus]
  {10.1016/j.icarus.2013.06.027}, \href
  {https://ui.adsabs.harvard.edu/abs/2013Icar..226..723D} {226, 723}

\bibitem[\protect\citeauthoryear{{DeMeo}, {Binzel}, {Slivan}  \& {Bus}}{{DeMeo}
  et~al.}{2009}]{DeMeo+2009}
{DeMeo} F.~E.,  {Binzel} R.~P.,  {Slivan} S.~M.,   {Bus} S.~J.,  2009, \mn@doi
  [\icarus] {10.1016/j.icarus.2009.02.005}, \href
  {https://ui.adsabs.harvard.edu/abs/2009Icar..202..160D} {202, 160}

\bibitem[\protect\citeauthoryear{{Geake} \& {Dollfus}}{{Geake} \&
  {Dollfus}}{1986}]{Geake+1986}
{Geake} J.~E.,  {Dollfus} A.,  1986, \mn@doi [\mnras] {10.1093/mnras/218.1.75},
  \href {https://ui.adsabs.harvard.edu/abs/1986MNRAS.218...75G} {218, 75}

\bibitem[\protect\citeauthoryear{{Geem} et~al.,}{{Geem}
  et~al.}{2022}]{Geem+2022}
{Geem} J.,  et~al., 2022, \mn@doi [\mnras] {10.1093/mnrasl/slac072}, \href
  {https://ui.adsabs.harvard.edu/abs/2022MNRAS.516L..53G} {516, L53}

\bibitem[\protect\citeauthoryear{Hirabayashi et~al.,}{Hirabayashi
  et~al.}{2021}]{Hirabayashi+2021}
Hirabayashi M.,  et~al., 2021, \mn@doi [Advances in Space Research]
  {https://doi.org/10.1016/j.asr.2021.03.030}, 68, 1533

\bibitem[\protect\citeauthoryear{{Ishiguro}, {Nakayama}, {Kogachi}, {Mukai},
  {Nakamura}, {Hirata}  \& {Okazaki}}{{Ishiguro} et~al.}{1997}]{Ishiguro+1997}
{Ishiguro} M.,  {Nakayama} H.,  {Kogachi} M.,  {Mukai} T.,  {Nakamura} R.,
  {Hirata} R.,   {Okazaki} A.,  1997, \mn@doi [\pasj] {10.1093/pasj/49.5.L31},
  \href {https://ui.adsabs.harvard.edu/abs/1997PASJ...49L..31I} {49, L31}

\bibitem[\protect\citeauthoryear{{Ishiguro} et~al.,}{{Ishiguro}
  et~al.}{2017}]{Ishiguro+2017}
{Ishiguro} M.,  et~al., 2017, \mn@doi [\aj] {10.3847/1538-3881/aa8b1a}, \href
  {https://ui.adsabs.harvard.edu/abs/2017AJ....154..180I} {154, 180}

\bibitem[\protect\citeauthoryear{{Ishiguro} et~al.,}{{Ishiguro}
  et~al.}{2022}]{Ishiguro+2022}
{Ishiguro} M.,  et~al., 2022, \mn@doi [\mnras] {10.1093/mnras/stab3198}, \href
  {https://ui.adsabs.harvard.edu/abs/2022MNRAS.509.4128I} {509, 4128}

\bibitem[\protect\citeauthoryear{{Kuroda}, {Ishiguro}, {Naito}, {Watanabe},
  {Hasegawa}, {Takagi}  \& {Kuramoto}}{{Kuroda} et~al.}{2021}]{Kuroda+2021}
{Kuroda} D.,  {Ishiguro} M.,  {Naito} H.,  {Watanabe} M.,  {Hasegawa} S.,
  {Takagi} S.,   {Kuramoto} K.,  2021, \mn@doi [\aap]
  {10.1051/0004-6361/202039004}, \href
  {https://ui.adsabs.harvard.edu/abs/2021A&A...646A..51K} {646, A51}

\bibitem[\protect\citeauthoryear{Lazzarin, Marchi, Magrin  \&
  Licandro}{Lazzarin et~al.}{2005}]{Lazzarin+2005}
Lazzarin M.,  Marchi S.,  Magrin S.,   Licandro J.,  2005, \mn@doi [Monthly
  Notices of the Royal Astronomical Society]
  {10.1111/j.1365-2966.2005.09006.x}, 359, 1575

\bibitem[\protect\citeauthoryear{{Lee} \& {Ishiguro}}{{Lee} \&
  {Ishiguro}}{2018}]{Lee2018}
{Lee} M.,  {Ishiguro} M.,  2018, \mn@doi [\aap] {10.1051/0004-6361/201832721},
  \href {https://ui.adsabs.harvard.edu/abs/2018A&A...616A.178L} {616, A178}

\bibitem[\protect\citeauthoryear{{Lumme} \& {Muinonen}}{{Lumme} \&
  {Muinonen}}{1993}]{Lumme+1993}
{Lumme} K.,  {Muinonen} K.,  1993, \mn@doi [IAU Symp.160: Asteroids, Comets,
  Meteors 1993] {10.1016/j.icarus.2016.11.003}, \href
  {http://adsabs.harvard.edu/abs/1993LPICo.810..194L} {160, 194}

\bibitem[\protect\citeauthoryear{Lupishko}{Lupishko}{2018}]{Lupishko+2018}
Lupishko D.~F.,  2018, \mn@doi [Solar System Research]
  {10.1134/S0038094618010069}, 52, 98

\bibitem[\protect\citeauthoryear{{Lupishko}}{{Lupishko}}{2022}]{Lupishko+2022}
{Lupishko} D.,  2022, \mn@doi [NASA Planetary Data System]
  {10.26033/hyf9-4762}, \href
  {https://ui.adsabs.harvard.edu/abs/2022pdss.data....1L} {p.~1}

\bibitem[\protect\citeauthoryear{Marsset et~al.,}{Marsset
  et~al.}{2020}]{Marsset+2020}
Marsset M.,  et~al., 2020, \mn@doi [The Astrophysical Journal Supplement
  Series] {10.3847/1538-4365/ab7b5f}, 247, 73

\bibitem[\protect\citeauthoryear{{Salvatier}, {Wiecki{\^a}}  \&
  {Fonnesbeck}}{{Salvatier} et~al.}{2016}]{Salvatier+2016}
{Salvatier} J.,  {Wiecki{\^a}} T.~V.,   {Fonnesbeck} C.,  2016, {Astrophysics
  Source Code Library, record ascl:1610.016}

\bibitem[\protect\citeauthoryear{{Schmidt}, {Elston}  \& {Lupie}}{{Schmidt}
  et~al.}{1992}]{Schmidt+1992}
{Schmidt} G.~D.,  {Elston} R.,   {Lupie} O.~L.,  1992, \mn@doi [\aj]
  {10.1086/116341}, \href
  {https://ui.adsabs.harvard.edu/abs/1992AJ....104.1563S} {104, 1563}

\bibitem[\protect\citeauthoryear{{Shkuratov} \& {Opanasenko}}{{Shkuratov} \&
  {Opanasenko}}{1992}]{Shkuratov+1992}
{Shkuratov} I.~G.,  {Opanasenko} N.~V.,  1992, \mn@doi [\icarus]
  {10.1016/0019-1035(92)90161-Y}, \href
  {https://ui.adsabs.harvard.edu/abs/1992Icar...99..468S} {99, 468}

\bibitem[\protect\citeauthoryear{{Tatsumi} et~al.,}{{Tatsumi}
  et~al.}{2018}]{Tatsumi2018}
{Tatsumi} E.,  et~al., 2018, \mn@doi [\icarus] {10.1016/j.icarus.2018.04.001},
  \href {https://ui.adsabs.harvard.edu/abs/2018Icar..311..175T} {311, 175}

\bibitem[\protect\citeauthoryear{{Turnshek}, {Bohlin}, {Williamson}, {Lupie},
  {Koornneef}  \& {Morgan}}{{Turnshek} et~al.}{1990}]{Turnshek+1990}
{Turnshek} D.~A.,  {Bohlin} R.~C.,  {Williamson} R.~L. I.,  {Lupie} O.~L.,
  {Koornneef} J.,   {Morgan} D.~H.,  1990, \mn@doi [\aj] {10.1086/115413},
  \href {https://ui.adsabs.harvard.edu/abs/1990AJ.....99.1243T} {99, 1243}

\bibitem[\protect\citeauthoryear{{Watanabe}, {Takahashi}, {Sato}, {Watanabe},
  {Fukuhara}, {Hamamoto}  \& {Ozaki}}{{Watanabe} et~al.}{2012}]{Watanabe+2012}
{Watanabe} M.,  {Takahashi} Y.,  {Sato} M.,  {Watanabe} S.,  {Fukuhara} T.,
  {Hamamoto} K.,   {Ozaki} A.,  2012, in Ground-based and Airborne
  Instrumentation for Astronomy IV. p. 84462O, \mn@doi{10.1117/12.925292}

\bibitem[\protect\citeauthoryear{{Whittet}, {Martin}, {Hough}, {Rouse},
  {Bailey}  \& {Axon}}{{Whittet} et~al.}{1992}]{Whittet+1992}
{Whittet} D.~C.~B.,  {Martin} P.~G.,  {Hough} J.~H.,  {Rouse} M.~F.,  {Bailey}
  J.~A.,   {Axon} D.~J.,  1992, \mn@doi [\apj] {10.1086/171039}, \href
  {https://ui.adsabs.harvard.edu/abs/1992ApJ...386..562W} {386, 562}

\bibitem[\protect\citeauthoryear{{Wolff}, {Nordsieck}  \& {Nook}}{{Wolff}
  et~al.}{1996}]{Wolff+1996}
{Wolff} M.~J.,  {Nordsieck} K.~H.,   {Nook} M.~A.,  1996, \mn@doi [\aj]
  {10.1086/117833}, \href
  {https://ui.adsabs.harvard.edu/abs/1996AJ....111..856W} {111, 856}

\makeatother
\end{thebibliography}





\bsp	
\label{lastpage}
\end{document}